\documentstyle[12pt]{article}

\begin{document}
{\sf \begin{center} \noindent {\Large \bf Differential rotation of stretched and twisted thick magnetic flux tube dynamos in Riemannian spaces}\\[3mm]

by \\[0.3cm]{\sl L.C. Garcia de Andrade}\\

\vspace{0.5cm} Departamento de F\'{\i}sica
Te\'orica -- IF -- Universidade do Estado do Rio de Janeiro-UERJ\\[-3mm]
Rua S\~ao Francisco Xavier, 524\\[-3mm]
Cep 20550-003, Maracan\~a, Rio de Janeiro, RJ, Brasil\\[-3mm]
Electronic mail address: garcia@dft.if.uerj.br\\[-3mm]
\vspace{2cm} {\bf Abstract}
\end{center}
\paragraph*{}
The topological mapping between a torus of big radius and a sphere
is applied to the Riemannian geometry of a stretched and twisted
very thick magnetic flux tube, to obtain spherical dynamos solving
the magnetohydrodynamics (MHD) self-induction equation for the
magnetic flux tubes undergoing differential (non-uniform) rotation
along the tube magnetic axis. Constraints on the shear is also
computed. It is shown that when the hypothesis of the convective
cyclonic dynamo is used the rotation is constant and a solid
rotational body is obtained. As usual toroidal fields are obtained
from poloidal magnetic field and these fields may be expressed in
terms of the differential rotation ${\Omega}(r,{\theta}(s))$. In the
case of non-cyclonic dynamos the torsion in the Frenet frame is
compute in terms of the dynamo constant. The flux tube shear
$\frac{\partial}{{\partial}r}{\Omega}$ is also computed. The
untwisted tube case is shown to be trivial in the sense that does
not support any dynamo action. This case is in agreement with
Cowling큦 antidynamo theorem, since in the untwisted case the tube
becomes axially symmetric which the refereed theorem rules out. We
also show that it is consistent with the Zeldovich antidynamo
theorem which rules out planar dynamos. Knowledge of the
differential rotation of the Earth, for example, allows one to place
limits on the curvature and torsion of the flux tube axis and
vice-versa , knowledge of the topology permit us to infer
differential rotation and other physical parameters of the stars and
planets. \vspace{0.5cm} \noindent {\bf PACS numbers:}
\hfill\parbox[t]{13.5cm}{02.40.Hw-Riemannian geometries}

\newpage
\section{Introduction}
 Earlier Parker \cite{1} investigated the so-called convective cell cyclonic dynamos in terms of the differential
 rotation of celestial and astrophysical objects and  solutions of self-induction equation
 equations coupled with the self-induction equation of the magnetic fields. Non-homogeneous rotation of celestial and
 astrophysical \cite{2} also called differential rotation, happens due to the fact that
 these bodies are not solid, but undergo distinct rotations between
 the poles and equator. This physical phenomena is supposed to
 produce and amplify magnetic fields in the so-called dynamo
 mechanics. On the other hand, Arnold, Zeldovich, Ruzmaikin and Sokoloff \cite{3} showed dynamos could considered
 as stationary solutions of self-induction equations in Riemannian three-dimensional spaces. Here we consider the Riemann
 metric of a very thick, stretched and twisted magnetic flux tube recently Garcia de Andrade \cite{4,5} to
 investigate magnetic flux tubes in superconducting plasmas, and use the map between spheres and very thich tori to obtain
 spherical dynamo solutions of MHD self-induction equation. Thiffault and Boozer \cite{6} following the same reasoning applied
 the methods of Riemann geometry in the context of chaotic flows and fast dynamos. Yet more recently Thiffeault \cite{7}
 investigated the stretching and Riemannian curvature of material lines in chaotic flows as possible
 dynamos models. In this paper he argued that filaments
 with torsion can also be constructed. Also Boozer \cite{8} has obtained a geomagnetic dynamo from conservation of magnetic helicity. This can also be shown here in the
 generalization to non-holonomic Frenet frame \cite{9}. Since in the case of kinematical dynamos we address here, the flow
 does not depend on the magnetic field (nonlinear MHD), we consider that the differential rotation ${\Omega}(r,{\theta}(s))$ depends
 on the radial and angular coordinates.  Since we
 know that the Zeldovich \cite{10} stretch, twist and fold method allows us to obtain kinematical dynamos, the stretch and twist of the
 magnetic flux tube seems providential to be consider as the germ of a spherical dynamo which has been so usually employed
 to explain the origin and physical nature of the geomagnetic and solar magnetic fields. The paper is organised as follows:
 In section II a brief outline of the explanation of the Riemannian geometry of magnetic flux tubes and how we may
 transform it into the Riemann metric \cite{11} of a sphere is presented. Section III presents anti-dynamo tests
 theorems of Zeldovich and Cowling큦 to show that both of them are fulfilled by the stretched, and twisted thick flux tube. In
 section IV general thick flux tube solutions of the self-induction equation on the Riemannian magnetic flux tube background metric as is usually done in handling Maxwell큦
 equations in Einstein큦 general theory of relativity, are presented. In section V conclusions are presented.
 \section{Spherical dynamos from closed flux tubes}
 According to Ricca \cite{12} the Riemann metric of a twisted magnetic flux tube
 may be written as
\begin{equation}
{dl}^{2}=dr^{2}+r^{2}d{{\theta}_{R}}^{2}+{K^{2}}(s)ds^{2} \label{1}
\end{equation}
where the tube coordinates are $(r,{\theta}_{R},s)$ \cite{12} where
${\theta}(s)={\theta}_{R}-\int{{\tau}ds}$ where $\tau$ is the Frenet
torsion and $\kappa$ is the curvature of the tube axis and $K(s)$ is
given by
\begin{equation}
{K^{2}}(s)=[1-r{\kappa}(s)cos{\theta}(s)]^{2} \label{2}
\end{equation}
Note that the limit of a very thin tube is $K:=1$, since the radial
coordinate r tends to zero by shrinkink the tube. But if we
substitute this value of K into the flux tube Riemann metric
(\ref{1}) one obtains
\begin{equation}
{dl}^{2}=dr^{2}+r^{2}d{{\theta}_{R}}^{2}+ds^{2} \label{3}
\end{equation}
which by substituting the coordinate-s along the magnetic flux tube
axis by the straight cylindrical coordinate-z one obtains the
Riemann flat metric of a very thin cylindrical tube, which implies
that when we compute the self-induced equations in this metric the
presence of the tube will not be fully felt by this equation. To
remedy this situation, we shall address here the other extreme of a
very thick tube where the internal radius can even surpass the value
of the radius of the tube. In this case expression (\ref{2}) becomes
\begin{equation}
{K^{2}}(s)=[r{\kappa}(s)cos{\theta}(s)]^{2} \label{4}
\end{equation}
by taking into account that the Frenet curvature
${\kappa}=\frac{1}{R}$ where R is the local radius of the tube,
substitution of (\ref{4}) into (\ref{1}) yields
\begin{equation}
{dl}^{2}=dr^{2}+r^{2}[d{{\theta}_{R}}^{2}+cos^{2}{\theta}(s)d{\phi}^{2}]\label{5}
\end{equation}
where the coordinate ${\phi}$ given by $d{\phi}:=\frac{ds}{R}$ is
the angular coordinate  in the plane of the torus. Expression
(\ref{5}) is clearly the Riemann metric describing the 3D sphere.
Since most of the celestial bodies including planets and stars can
be described by spherical or spheroidal symmetries, solving the
self-induction equation for non-turbulent fluids inside the twisted
and stretched thick magnetic flux tubes, seems to be the a useful
approach to dynamo theories. Computing the Riemannian gradient
operator ${\nabla}$ in terms of the thick flux tube curvilinear
coordinates \cite{13} reads
\begin{equation}
{\nabla}=[{\partial}_{r},\frac{1}{r}{\partial}_{\theta},\frac{1}{K}{\partial}_{s}]
\label{6}
\end{equation}
where ${\partial}_{j}:=\frac{{\partial}}{{\partial}x^{j}}$. The
magnetic field here can be expressed as
\begin{equation}
\vec{B}=e^{pt}\vec{B}_{0}:=e^{pt}[B_{\theta}(r,{\theta})\vec{e_{\theta}}+B_{s}(r)\vec{t}]
\label{7}
\end{equation}
where p is the dynamo constant and real parameter here, the dynamo
condition is $p\ge0$. These solutions will be tested as dynamos in
the next section.

\section{Testing Cowling and Zeldovich anti-dynamo theorems in stretched and twisted thick flux tubes}
To test Cowling큦 antidynamo theorem which states that axially
symmetric magnetic devices cannot support dynamo action, in this
section we solve the MHD equations
\begin{equation}
{\nabla}.\vec{B}=0 \label{8}
\end{equation}
\begin{equation}
\frac{{\partial}}{{\partial}t}\vec{B}-{\nabla}{\times}[\vec{v}{\times}\vec{B}]-{\epsilon}{\nabla}^{2}\vec{B}=0
 \label{9}
\end{equation}
\begin{equation}
{\nabla}.\vec{v}=0 \label{10}
\end{equation}
where $\vec{u}$ is a solenoidal field while ${\epsilon}$ is the
resistivity coefficient. Equation (\ref{8}) represents the induction
equation. The expression $\vec{v}=[0,{\Omega}r,v_{0}]$ where $v_{0}$
is the constant speed of the flow along the magnetic axis. Since
here we shall only consider nondissipative flows , ${\epsilon}$
vanishes and we do not need to compute the Riemannian Laplacian
${\nabla}^{2}$. Here
\begin{equation}
\vec{e_{\theta}}=-\vec{n}cos{\theta}+\vec{b}sin{\theta} \label{11}
\end{equation}
which by using the Frenet frame relations
\begin{equation}
\vec{t}'=\kappa\vec{n} \label{12}
\end{equation}
\begin{equation}
\vec{n}'=-\kappa\vec{t}+ {\tau}\vec{b} \label{13}
\end{equation}
\begin{equation}
\vec{b}'=-{\tau}\vec{n} \label{14}
\end{equation}
where the dash represents the ordinary derivation with respect to
coordinate-s , yields
\begin{equation}
\frac{\partial}{{\partial}s}\vec{e_{\theta}}={\kappa}sin{\theta}\vec{t}
\label{15}
\end{equation}
Substitution of these expressions into the MHD equations
yields
\begin{equation}
p\vec{B}_{0}+[\frac{v_{\theta}}{r}{\partial}_{\theta}+\frac{v_{0}}{K}{\partial}_{s}]\vec{{B}_{0}}-[\frac{B_{\theta}}{r}{\partial}_{\theta}+\frac{B_{s}}{K}{\partial}_{s}]\vec{v}=0
\label{16}
\end{equation}
Substitution of expressions (\ref{7}), (\ref{11}) and (\ref{15})
into (\ref{16}) yields the following three scalar equations along
the Frenet basis $(\vec{t},\vec{n},\vec{b})$
\begin{equation}
p={\tau}tg{\theta}[v_{0}+\frac{{\Omega}r}{Tw}] \label{17}
\end{equation}
\begin{equation}
p=\frac{\tau}{Tw}[-v_{0}+{\Omega}r]\label{18}
\end{equation}
\begin{equation}
p=\frac{v_{0}tg{\theta}}{r}[1+{Tw}] \label{19}
\end{equation}
Here we consider the twist definition as
$Tw=\frac{B_{\theta}}{B_{s}}$.We also consider in this derivation
the other MHD equations
\begin{equation}
\frac{{\partial}B_{\theta}}{{\partial}s}=\frac{B_{\theta}{\kappa}{\tau}rsin{\theta}}{K}
\label{20}
\end{equation}
which is valid also for $v_{\theta}$. An immediate astrophysical
consequence of these equations is that the twisted flux tube does
not support dynamo action when the tube is planar. By planar here,
we mean that the torsion of the magnetic axis is planar which
geometrically means that the Frenet torsion vanishes. This is
exactly the Zeldovich anti-dynamo theorem \cite{10}. To prove this
result here we make the substitution ${\tau}=0$ into equations
(\ref{17}),(\ref{18}), and (\ref{19}), which yields
\begin{equation}
p=0 \label{21}
\end{equation}
and
\begin{equation}
 p=\frac{v_{0}tg{\theta}}{r}[1+Tw]\label{22}
\end{equation}
these two last equations together imply that either $v_{0}=0$ or
coordinate r at infinity. The first and more realistic hypothesis
yield a planar circular flow. In other words, the helical flux
inside the magnetic flux tube reduces to a planar circular flow. Of
course our main result is in equation (\ref{21}) which shows the
dynamo action is not supported. Let us now turn our attention to
show that the Cowling theorem is fulfilled, or that the untwisted,
axially symmetric flux tubes do not support dynamo actions. To
accomplished this task we simply substitute the expression $Tw=0$
for the unwisted tube, into the same equations were used to test
Zeldovich theorem, which in turn yields
\begin{equation}
{\tau}tg{\theta}{\Omega}r=0 \label{23}
\end{equation}
\begin{equation}
v_{0}={\Omega}r\label{24}
\end{equation}
\begin{equation}
p=\frac{v_{0}tg{\theta}}{r} \label{25}
\end{equation}
Assuming that the torsion does not vanish in the equation
(\ref{23}), and since the tube being thick, coordinate r cannot
vanish, we conclude that the angular velocity ${\Omega}$ of the flow
also vanish which from equation (\ref{24}) that $v_{0}$ and from the
last equation (\ref{25}) we obtain $p=0$. In the next section we
shall analyse the general dynamo solution of the stretched , twisted
tube.
\section{Differential rotation of stretched, twisted thick flux tubes
dynamos}
In this section we shall consider the general solution of
self-induced equations and also consider the constraints of the
other divergence-free equations on the non-uniform motion
(differential rotation) of the flux tube. Expressions
(\ref{17}),(\ref{18}) and (\ref{19}) altogether yields an algebraic
equation to the twist of the tube, given by
\begin{equation}
Tw^{2}-[{\tau}r-v_{0}]Tw-\frac{{\tau}{\Omega}r^{2}}{v_{0}}=0
\label{26}
\end{equation}
Solutions of this algebraic equation for the tube twist are
\begin{equation}
Tw=\frac{1}{2}[{\tau}r\pm(({\tau}r)^{2}[1+4{\Omega}])^{\frac{1}{2}}]
\label{27}
\end{equation}
to simplify this first solution let us assume the cyclonic
hypothesis \cite{1}, where ${\Omega}>>1$. Substitution of this value
into the last expression yields
\begin{equation}
Tw=\frac{1}{2}{\tau}r[\pm(2{\Omega})^{\frac{1}{2}}] \label{28}
\end{equation}
where we have consider the strong torsion bound ${\tau}r>>v_{0}$.
This allows us to determine the differential rotation in terms of
the twist as
\begin{equation}
\frac{1}{2}{\Omega}(r,{\theta}(s))=\frac{Tw^{2}}{{\tau}^{2}r^{2}}
\label{29}
\end{equation}
Since the flux tube twist is given by the ratio between the poloidal
and toroidal components of the magnetic field, we obtain a relation
between these components and the differential rotation as
\begin{equation}
{B_{s}}^{2}=\frac{\sqrt{2}}{2}\frac{{\Omega}}{[{\tau}r]^{2}}{B_{\theta}}^{2}
\label{30}
\end{equation}
Assuming that the cyclonic hypothesis also implies that
${\Omega}>>v_{0}$ equation (\ref{18}) becomes
\begin{equation}
p=\frac{\tau}{Tw}[{\Omega}r]\label{31}
\end{equation}
Substitution of expression (\ref{29}) into last expression yields
\begin{equation}
p=\sqrt{2{\Omega}}>0\label{32}
\end{equation}
Since $p>0$ a dynamo action is supported , however since p by
hypothesis is constant, the differential rotation degenerates in a
solid homogeneous rotation. Let us now investigate the case of
non-cyclonic rotation where ${\Omega}<<1$ and ${\Omega}<<v_{0}$.
Under these bounds the twist algebraic solutions reduces to
$Tw={\tau}r$. Substitution of this result into the expression
(\ref{18}) yields
\begin{equation}
p={\Omega}\label{33}
\end{equation}
which supports also anti-dynamos or non-dynamos (p<0) for
anti-cyclonic rotations $({\Omega}<0)$. To further investigate the
diferential rotation let us consider the equation
\begin{equation}
\frac{{\partial}v_{\theta}}{{\partial}s}=\frac{v_{\theta}{\kappa}{\tau}rsin{\theta}}{K}
\label{34}
\end{equation}
which by substitution of the thickness condition on K and
$v_{\theta}={\Omega}r$ yields the shear relation
\begin{equation}
\frac{{\partial}{\Omega}}{{\partial}s}={\Omega}{\tau}tg{\theta}
\label{35}
\end{equation}
Since in this our case ${\Omega}$ is constant either ${\Omega}$ or
$\tau$ vanishes, which does not represent dynamos as we have just
seen, or yet $tg{\theta}$ vanishes which yields ${\theta}=0$ region.
The only dynamo condition which finally survives is to consider that
the product between torsion and ${\Omega}=p$ product is very weak.
This is however a not very efficient dynamo since though $p>0$ , it
is close to zero, which also gives a very weak rotation which also
yields an anti-dynamo as for example, in the case of the Venus
planet which does not support a magnetic field provenient from
dynamo actions since its rotation is 243 lower that the Earth큦. As
a final attempt to obtain our dynamo solution let us drop the strong
torsion of the flux tube dynamo and ${\tau}r=v_{0}$, which from
expression (\ref{26}) yields
\begin{equation}
Tw^{2}={\Omega}r \label{36}
\end{equation}
which upon substitution into expression (\ref{18}) yields the
differential rotation as
\begin{equation}
{\Omega}(r,s)=\frac{p^{2}}{{\tau}^{2}r} \label{37}
\end{equation}
which is now is not constant and a true differential rotation and
besides it is also a dynamo since $p={\tau}\sqrt{{\Omega}r}>0$ if
${\tau}>0$. The shear also does not vanish and is written as
$\frac{{\partial}{\Omega}}{{\partial}s}=-\frac{2p^{2}}{r}{\tau}^{-3}tg{\theta}$
since ${\tau}>0$. Substitution of this last result into equation
(\ref{35}), allows us to determine the torsion in terms of the
dynamo constant p as  ${\tau}=\frac{p}{\sqrt{rtg{\theta}}}$ which to
be real only on certain branches of the flux tube unless the
constant p or this purely imaginary, which gives us the general
dynamo condition as $Re(p)>0$ \cite{14}.
\section{Conclusions}
 In conclusion, we have tested Cowling and Zeldovich anti-dynamo theorems in thick stretched and twisted flux tubes, as
 solution of MHD cyclonic flows. Two dynamo solutions are obtained , one which is a very inefficient dynamo and the other which
 is a better efficient dynamo where the differential rotation is also computed along the shear along the magnetic axis of the tube.
 Since the Riemann metric of the very thick tube coincides with the sphere metric in three dimensions we may argue that
 spherical dynamos are also obtained by this technique. Other interesting test for Cowling antidynamo theorem can be obtained
 by using other metrics besides the four-dimensional black hole spacetime considering recently numerically by
 Brandenburg \cite{15}. Future prospects
 included the investigation of general relativistic MHD dynamos on
 the background of Lewis metric.
 \section*{Acknowledgements}
 Thanks are due to Professor King Hay Tsui for helpful discussions on the subject of this paper, and to CNPq and UERJ for financial supports.


\newpage
\begin{thebibliography}{15}
\bibitem{1} E.N. Parker, Cosmical Magnetic Fields (1980), Oxford University Press.
\bibitem{2} G. Rudiger and R. Hollerbach, The Magnetic Universe
(2004), Wiley.
\bibitem{3} V. Arnold, Ya B. Zeldovich, R. Ruzmaikin, D. Sokoloff, Soviet Physics J. JETP 54 (6) , (1981) 1083.
\bibitem{4} L. C. Garcia de Andrade, Physics of Plasmas 13, 022309 (2006).
\bibitem{5} L.C. Garcia de Andrade, Twist transport in strongly
torsioned astrophysical flux tubes,Astrophysics and Space Science
(2007) in press.
\bibitem{6} J. Thiffeault and A.H.Boozer,Chaos 11,(2001) 16.
\bibitem{7} J. Thiffeault, Stretching and Curvature of Material
Lines in Chaotic Flows,(2004) Los Alamos arXiv:nlin. CD/0204069.
\bibitem{8} A.H. Boozer, Phys. of Fluids B 5 (7), (1993) 2271.
\bibitem{9} C. Thakur, Astrophysics and space science 149 (1988) 83.
\bibitem{10} Ya B. Zeldovich, A.A. Ruzmaikin and D.D. Sokoloff, The
Almighty Chance (1990) World sci. Press.
S. Childress and A.
Gilbert, Stretch, Twist and Fold: The Fast Dynamo (1996)(Springer).
\bibitem{11} V. Arnold and B. Khesin, Topological Methods in Hydrodynamics,
Applied Mathematics Sciences 125 (1991) Springer.
\bibitem{12} R. Ricca, Solar Physics 172,241 (1997).
\bibitem{13} W.D. D'haesseleer, W. Hitchon, J. Callen and J.L. Shohet, Flux Coordinates and Magnetic field Structure (1991) Spinger.
\bibitem{14} S. Childress and A.
Gilbert, Stretch, Twist and Fold: The Fast Dynamo (1996)(Springer).
\bibitem{15} A. Brandenburg, Ap. J 465:L115 (1996).
\end{thebibliography}
\end{document}